\documentstyle[12pt,titlepage]{article}
\newlength{\extraspace}
\newlength{\extraspaces}
\catcode`\@=11
 
\def\numberbysection{\@addtoreset{equation}{section}
\def\theequation{\arabic{section}.\arabic{equation}}}

\newcommand{\be}{\begin{equation}
\addtolength{\abovedisplayskip}{\extraspaces}
\addtolength{\belowdisplayskip}{\extraspaces}
\addtolength{\abovedisplayshortskip}{\extraspace}
\addtolength{\belowdisplayshortskip}{\extraspace}}
\newcommand{\ee}{\end{equation}}
\newcommand{\ba}{\begin{eqnarray}
\addtolength{\abovedisplayskip}{\extraspaces}
\addtolength{\belowdisplayskip}{\extraspaces}
\addtolength{\abovedisplayshortskip}{\extraspace}
\addtolength{\belowdisplayshortskip}{\extraspace}}
\newcommand{\ea}{\end{eqnarray}}

\newcommand{\tr}{\, {\rm tr} \,}

\begin{document}
\begin{titlepage}
\addtolength{\baselineskip}{.7mm}
\thispagestyle{empty}
%
%
\begin{center}
{\large{\bf Three loop renormalization group \\ for
a marginally perturbed SU(2) WZW model}} \\[15mm]
\vspace*{3cm}
{\sc Chigak Itoi}
\footnote{\tt e-mail: itoi@phys.cst.nihon-u.ac.jp} \\[2mm]
{\it Department of Physics, \\[1mm]
College of Science and Technology, Nihon University, \\[1mm]
Kanda Surugadai, Chiyoda, Tokyo 101, Japan} \\[6mm]
\vspace*{3cm}
{\bf Abstract}\\[5mm]
{\parbox{13.5cm}{\hspace{5mm}
Employing a simple calculation method obtained by M.-H. Kato, 
we calculate  
the three loop renormalization group
in the $su(2)$ coset conformal field theory
with a slightly relevant perturbation and
the $su(2)$ Wess-Zumino-Witten 
model with a particular invariant marginal
perturbation.  
Zamolodchikov's $c$-theorem,
exact data of the perturbation 
operator and a known exact form
of the operator product coefficient
enable us to calculate the beta function, 
the gamma function and
the $c$-function to three loop order. 
This result gives 
the logarithmic finite size correction 
to the ground state energy and 
the low temperature behavior of the specific heat in the 
Heisenberg antiferromagnetic chain with high accuracy.
We describe the consistency with results obtained by several authors
on the basis of its exact solvability.
We discuss an experiment 
of the specific heat and the suceptibility 
recently observed.}}
\end{center}
{\it keywords}: c-function, renormalization group, marginal operator,
three loop, logarithmic correction 
\vspace*{5ex}
\end{titlepage}
\setcounter{section}{0}
\setcounter{equation}{0}
%
\section{Introduction}
There have been extensive studies in conformal field theory (CFT) with
a relevant perturbation which reveal universal nature of 
critical phenomena in low
dimensional statistical physics.
Landau-Ginzburg theory gives a good qualitative picture
for the  renormalization group (RG) flow 
including multi-critical points \cite{LG}. 
In this theory, we can specify the driving relevant operator 
of the CFT at the ultraviolet (UV) fixed point 
and the irrelevant operator
of another CFT at the infrared (IR) fixed point.
We can check the expected flow practically 
in the $\varepsilon$-expansion by calculating
 the deviation of the central charge which 
obeys Zamolodchikov's $c$-theorem \cite{Z}. 

In this paper, we utilize
Zamolodchikov's $c$-theorem and 
the exact form of the operator product expansion (OPE) coefficient
to calculate the beta function
to three loop order in the $\varepsilon$-expansion
for $su(2)$ coset models with the 
slightly relevant perturbation with the dimension $2-\varepsilon$. 
M. -H. Kato argued this application of the $c$-theorem to
calculate the beta function up to two loop order 
for the slightly perturbed general coset models
in all $A$-$D$-$E$ classes \cite{K}. 
By taking the limit $\varepsilon \rightarrow 0$,
the coset CFT becomes a certain Wess-Zumino-Witten (WZW) model 
with a particular invariant marginal perturbation. 
In this limit, he also calculated 
the logarithmic finite size correction
to the ground state energy and the logarithmic 
temperature dependence of the 
specific heat of quantum spin chains
to two loop order. The obtained fitting function
agrees with the groud state energy of an $su(2)$ spin chain 
by a numerical Bethe ansatz \cite{N}.
Here, we calculate higher order logarithmic correction to the 
ground state energy and  
discuss the consistency with results obtained by several
authors in different methods.
This paper is organized as follows. In section 2, we review 
M.-H. Kato's method to calculate two loop beta function
in general coset models
with generalizing some equations for 
higher order calculation. In section 3, we confine our discussions to 
$su(2)$ case. In $su(2)$ coset model, the known exact form of 
the OPE coefficient of the driving operator enables us to calculate 
the three loop beta function. In section 4, we discuss the 
logarithmic finite size correction to low lying energy levels.
We discuss consistency with a numerical Bethe ansatz and some other methods. 
In section 5, we calculate a logarithmic temperature dependence
of the specific heat per unit length in a $su(2)$ spin chain.
We discuss the recent experiment \cite{U} from the view point of
the higher order renormalization group calculation.
 
\section{Two loop calculation in general coset models}
Here, we review M.-H. Kato's method for two loop beta function
\cite{K} with preparing some extended forms of the equations 
for three loop calculation in the next section.
In this method, we consider perturbed coset CFT ${\cal M}(k,l; G)=
\hat{G}_k \oplus \hat{G}_l  /  \hat{G}_{k+l}$
by a slightly relevant operator $\Phi_{\rm UV}$ with self-closing algebra
which corresponds to the (1,3) operator in the Virasoro minimal model
as a regularized theory for the marginally perturbed WZW model. 
In this section, we assume those following two facts obtained by
Ahn, Bernard and Leclair \cite{ABL}.
They showed that the slightly relevant  
operator $\Phi_{\rm UV}$ drives the CFT at UV fixed point to
the IR one with the irrelevant perturbation
of the operator $\Phi_{\rm IR}$, as in $su(2)$
case \cite{LG,Z} 
\begin{equation}
H(k,l) + g \int \frac{d^2 z}{2 \pi} \Phi_{\rm UV}(z, \bar{z}) 
\rightarrow  H(k, l-k)
+ g' \int \frac{d^2 z }{2 \pi}\Phi_{\rm IR}(z \bar{z}), \ ({\rm IR \ limit}) 
\label{ass0}
\end{equation}
where $H(k,l)$ is a critical hamiltonian in ${\cal M}(k,l; G)$.  
They showed also that 
this deformed CFT ${\cal M}(k,l; G)$ by the operator $\Phi_{\rm UV}$ 
becomes the $\hat{G}_k$ WZW model with an invariant marginal perturbation,
that is
$$
\hat{G}_k \oplus \hat{G}_l / \hat{G}_{k+l} \rightarrow 
\hat{G}_k, \ (l \rightarrow \infty),
$$
and 
\begin{equation}
\Phi_{\rm UV}(z, \bar{z}) \rightarrow 
\Phi(z, \bar{z}) \equiv -\frac{2}{k \sqrt{D}} 
\sum_{a=1} ^D J^a(z) \bar{J}^a(\bar{z}), \ (l \rightarrow \infty),
\label{ass}
\end{equation}
in a limit $l \rightarrow \infty $,
where $D$ is the dimension of the 
Lie algebra $G$. 
The existence of RG flow 
between these fixed points is guaranteed by the 
self-closing algebra of the operator $\Phi_{\rm UV}$. 
In the CFT ${\cal M}(k,l; G)$, the central charge is
\begin{eqnarray}
c(k,l; G) &=& c(\hat{G}_k)+ c(\hat{G}_l)-c(\hat{G}_{k+l}) \\ \nonumber
&=& \frac{k D}{k+h_c} 
\left(1-\frac{h_c(k+h_c)}{(l+h_c)(k+l+h_c)} \right),
\end{eqnarray}
and the conformal dimensions of 
the operators $\Phi_{\rm UV}$ and  $\Phi_{\rm IR}$ are
\begin{eqnarray}
\Delta_{\rm UV}&=&1-\frac{h_c}{k+l+h_c} \\ \nonumber
\Delta_{\rm IR}&=&1+\frac{h_c}{l-k+h_c},
\end{eqnarray}
where $r$ is the rank and $h_c$ is the dual coxeter number of the 
Lie algebra $G$. 
The dimension of the Lie algebra is given by $D=r (1+h_c)$
in simply-laced algebra.
For example in $G = su(n)$, these are given by $r= n-1$, $h_c= n$
and $D=n^2-1$.
Here, 
we define the parameter 
\begin{equation}
\varepsilon \equiv 2-2 \Delta_{\rm UV} = \frac{2 h_c}{k+l+h_c},
\end{equation}
and discuss the RG flow from the UV
theory ${\cal M}(k,l;G)$ to 
the IR theory ${\cal M}(k,l-k; G)$ in the $\varepsilon$-expansion. 
The limit $\varepsilon \searrow 0$ as $l \nearrow \infty$
gives 
$$
\Delta_{\rm UV} = 1- \frac{\varepsilon}{2} \rightarrow 1.
$$
This relation is necessary for the identification
of the operator (\ref{ass}), which should be shown by the correspondence
between these OPE coefficients with their closing algebra,
as well as their conformal dimensions.
The OPE relations of the operator $\Phi_{\rm UV}$ 
and the marginal operator are
\begin{eqnarray}
\Phi_{\rm UV}(z, \bar{z}) \Phi_{\rm UV}(0, 0) &\sim& 
\frac{b(\varepsilon)}{|z|^{2 \Delta_{\rm UV}}} \Phi_{\rm UV}(0, 0) \\ 
\Phi(z, \bar{z}) \Phi(0, 0) &\sim& \frac{b_0}{|z|^2} \Phi(0, 0).
\label{corr}
\end{eqnarray}
with $b(\varepsilon) \rightarrow b_0$ as $\varepsilon \rightarrow 0$.
To obtain the OPE coefficient $b(\varepsilon)$ 
in the $\varepsilon$-expansion, we assume Zamolodchikov's $c$-theorem
\cite{Z}. 
The beta function of the deformed 
CFT by the operator $\Phi_{\rm UV}$ 
with the running coupling constant $g$ is
\begin{equation}
\beta(g) = -\varepsilon g + \frac{b(\varepsilon)}{2} g^2 + 
\frac{d(\varepsilon)}{2} g^3 + \frac{e(\varepsilon)}{2} g^4 + \cdots,
\label{beta}
\end{equation}
and the gamma function of the operator $\Phi_{\rm UV}$ is given by
\begin{equation}
\gamma(g) = 2 + \frac{\partial \beta(g)}{\partial g}.
\label{gamma}
\end{equation}
The coefficients in the beta function is expanded in $\varepsilon$
\begin{eqnarray}
b(\varepsilon) &=& \sum_{n=0} ^\infty  b_n \varepsilon ^n, \\ 
d(\varepsilon) &=& \sum_{n=0} ^\infty d_n \varepsilon ^n, \\ 
e(\varepsilon) &=& \sum_{n=0} ^\infty e_n \varepsilon ^n, \\ 
\cdots.
\label{coefbeta}
\end{eqnarray}
The beta function (\ref{beta}) has a trivial fixed point $g=0$ and 
another non-trivial one $g=g_* \neq 0$. The trivial fixed point $g=0$ 
is the UV CFT ${\cal M}(k,l; G)$ and the other one $g_*$ 
corresponds to the IR CFT ${\cal M}(k,l-k; G)$.
The non-trivial fixed point is expanded in a series
\begin{equation}
g_* = \sum_{n=1} ^\infty g_n \varepsilon^n, 
\end{equation}
whose coefficients $g_n$ are written in terms of those of the 
beta function (\ref{coefbeta})
\begin{eqnarray}
g_1 &=& \frac{2}{b_0}, \\ \nonumber
g_2 &=& -\frac{1}{b_0^3}\left(2 b_1 b_0^2 + 4 d_0 \right), \\ \nonumber
g_3 &=& \frac{2}{b_0^5}\left(b_0 ^2 b_1 ^2 -b_0 ^3 b_2 + 6 b_0 b_1 d_0
+ 8 d_0 ^2 -2 b_0 ^2 d_1 - 4 b_0 e_0 \right), \\ 
& & \cdots .
\end{eqnarray}
Zamolodchikov's $c$-function is defined by
\begin{equation}
c(g) = c_{\rm UV} + \frac{3}{2} \int_0 ^g \beta(x) dx,
\label{cfun}
\end{equation}
where $c_{\rm UV}$ is the central charge of the CFT ${\cal M}(k,l; G)$.
Zamolodchikov's $c$-theorem gives us the following constraints 
\begin{eqnarray}
c_{\rm IR} &=& c(g_*), \label{c1} \\
\Delta_{\rm IR} &=& \gamma(g_*)/2 = 1+ \beta'(g_*)/2, \label{c2}
\end{eqnarray}
where $c_{\rm IR}$ is the central charge of the IR CFT
${\cal M}(k,l-k; G)$.
The right hand side in these constraints can be expanded
in $\varepsilon$
\begin{equation}
c_{\rm IR}-c_{\rm UV} = -\frac{\varepsilon^3}{b_0^2} +
\left(\frac{2b_1}{b_0^3} + \frac{3d_0}{b_0^4}\right)
\varepsilon^4
+\left(\frac{2b_2}{b_0^3}+\frac{3d_1-3b_1^2}{b_0^4}+
\frac{-60b_1d_0+24e_0}{5b_0^5} -\frac{12d_0}{b_0^6}
\right) \varepsilon^5 + \cdots,
\label{c12}
\end{equation}
and 
\begin{equation}
\Delta_{\rm IR} = 1 + \frac{\varepsilon}{2} +
\frac{d_0}{b_0^2} \varepsilon^2
+ \frac{1}{b_0^4}\left(-2b_0b_1d_0-4d_0^2+b_0^2d_1
+4b_0e_0 \right) \varepsilon^3 + \cdots.
\label{c22}
\end{equation}
The knowledge of the flow ${\cal M}(k,l; G) \rightarrow {\cal M}(k,l-k ; G)$
by the slightly relevant perturbation gives further constraints
\begin{eqnarray}
c_{\rm IR} &=& c(k, l-k; G) = \frac{k D}{k+h_c}\left( 
1-\frac{h_c(k+h_c)}{(l-k+ h_c)(l+h_c)} \right),  \label{l1} \\
\Delta_{\rm IR} &=& 1+ \frac{h_c}{l-k+h_c}. \label{l2}
\end{eqnarray}
The $\varepsilon$-expansion of these constraints (\ref{l1}) and (\ref{l2})
can be done as follows:
\begin{eqnarray}
c_{\rm IR} - c_{\rm UV} &=& -\frac{k^2 D \varepsilon^3 }{4h_c ^2}
\left(1+\frac{3k}{2h_c}\varepsilon + \frac{7k^2}{4 h_c ^2} \varepsilon^2
+ \cdots \right),  \label{l12} \\
\Delta_{\rm IR} &=& 1+ \frac{\varepsilon}{2} +\frac{k}{2 h_c} \varepsilon^2
+ \frac{k^2}{2h_c ^2} \varepsilon^3 + \cdots. 
\label{l22}
\end{eqnarray}
These constraints (\ref{c12}), (\ref{c22}), (\ref{l12})
and (\ref{l22})
can be solved with respect to $b_1$ and $d_0$ in terms of $b_0$ 
\begin{equation}
b_1 = - \frac{3k}{2 h_c} b_0, \ \ \ \  d_0 =
\frac{k}{2h_c} b_0^2,
\end{equation}
where $b_0$ is OPE coefficient of the marginal operator in
the WZW model
$$
b_0=\frac{2h_c}{k\sqrt{D}}.
$$
Therefore, we obtain the beta function of the marginally
perturbed $\hat{G}_k$ WZW model up to two loop order
$$
\beta(g) = \frac{h_c}{k\sqrt{D}} g^2 +
\frac{h_c}{k D} g^3,
$$
without explicit calculation.

\section{Three loop beta function}

In the previous section, we obtained the two loop beta function
with the help of two assumptions (\ref{ass0}) and (\ref{ass}).
In this section, we limit our discussion 
to $G=su(2)$ case and we check these assumption within the
framework of the $\varepsilon$-expansion.
In this case, we have $D=3$, $h_c=2$ and $r=1$.
We have the slightly relevant 
operator $\Phi_{\rm UV}$ with the conformal dimension 
$\Delta_{\rm UV}=1-2/(k+l+2)$.
The exact form of its 
OPE coefficient $b(\varepsilon)$ in the CFT
${\cal M}(k,l; su(2))$ is given in \cite{CSS} 
\begin{eqnarray}
b(\varepsilon) &=& \frac{8}{\sqrt{3} k}
\frac{(2-(k+1)\varepsilon)^2}{(2-\varepsilon)(4-(k+2)\varepsilon)} \\
\nonumber
&\times& \frac{\Gamma(1-\varepsilon)}{\Gamma(1+\varepsilon)}
\left(\frac{\Gamma(1-\varepsilon/4)}{\Gamma(1+\varepsilon/4)}\right)^{3/2}
\left(\frac{\Gamma(1+3 \varepsilon/4)}{\Gamma(1-3
\varepsilon/4)}\right)^{1/2}
\left(\frac{\Gamma(1+\varepsilon/2)}{\Gamma(1-\varepsilon/2)}\right)^2.
\label{exb}
\end{eqnarray}
This gives us the data of $b_0, \  b_1, \ b_2, \cdots $ 
and then the method explained in the previous section
enables us to calculate the beta function up to third order.
The OPE coefficient (\ref{exb}) is expanded in 
\begin{equation}
b(\varepsilon) =\frac{4}{\sqrt{3} k}\left(1- \frac{3 k}{4}
\varepsilon +\frac{k(k-6)}{16} \varepsilon^2 + \cdots \right).
\end{equation}
This formula shows 
$$
b(\varepsilon) \rightarrow b_0 \equiv \frac{4}{\sqrt{3} k}
\ \ \ (\varepsilon \rightarrow 0),
$$
which justifies the identification (\ref{ass}) we assumed.
The one loop beta function is given by $\beta(g) = b_0 g^2 /2$
which enables us to calculate the central charge $c_{\rm IR}$
and the conformal dimension $\Delta_{\rm IR}$
by eq.(\ref{c12}) and eq.(\ref{c22}).
These results show that the expected flow (\ref{ass0}) is checked  
in the $\varepsilon$-expansion around the UV fixed point
at one loop level. 
Therefore, we can use the constrains to solve the
higher order coefficients in the $\varepsilon$-expansion.   
Since the constraints (\ref{c12}), (\ref{c22}),
(\ref{l12}) and (\ref{l22}) can be solved with respect to
$d_0,$ $d_1,$ and $e_0$ in terms of $b_0,$ $b_1,$ and
$b_2$,
we obtain them explicitly
\begin{equation}
d_0 = \frac{4}{3 k},  \ \ \ d_1 = -\frac{10(k-2)}{9k}, \ \
\  e_0 = \frac{10(k-2)}{9\sqrt{3} k^2},
\label{cobeta}
\end{equation}
where the rank and the dual coxeter number of $su(2)$
are given by $r=1$ and $h_c=2$.
The limit $\varepsilon \rightarrow 0$ gives the 
beta function to three loop order
for the marginally perturbed level $k$ $su(2)$ WZW
model 
\begin{equation}
\beta(g) = \frac{b_0}{2} g^2 + \frac{d_0}{2} g^3 +
\frac{e_0}{2} g^4 + {\rm O}(g^5).
\label{beta3} 
\end{equation}

\section{The $c$-function and the logarithmic correction}

Here, we consider the perturbed WZW 
model in the asymptotically non-free
region $g>0$. The marginally perturbed level $k$ $su(2)$ WZW model
in asymptotically non-free region
describes $S=k/2$ antiferromagnetic Faddeev-Takhtajan model. 
Especially for $k=1$, this corresponds to $S=1/2$
Heisenberg antiferromagnetic chain model.
We calculate the $c$-function by integrating the beta
function and discuss the logarithmic finite size
correction
to the ground state energy of the antiferromagnetic chain.
First, we integrate the differential equation 
\begin{equation}
\frac{dg}{d \log l} = - \beta(g)
\end{equation}
for the
running coupling
constant $g(l)$
\begin{equation}
I(g(l))-I(g(a)) = \log l/a,
\label{run}
\end{equation}
where $a$ is the UV cut off length scale of the field theory.
Here, we define the integral function $I(x)$ by
\begin{equation}
I(x) \equiv -\int \frac{dx}{\beta(x)} =
\frac{2}{b_0 x}+\frac{2d_0}{b_0^2} \log x -
\frac{2}{b_0^2}
\left(\frac{d_0^2}{b_0} - e_0 \right) x + {\rm O}(x^2).
\end{equation}
Let $L$ be the size of the chain which sufficiently large
compared to the UV cut off $a$. 
In this case, we have $g(L) << g(a)$
because of the asymptotic non-freedom. Here, we introduce
the RG invariant length scale $L_0 \equiv a \exp(-I(g(a)))$.
The running coupling constant $g(L)$ is determined by
$$
I(g(L))=\log L/L_0,
$$ 
and solve this equation by expanding
$1/I(g(L))$ 
in $g(L)$ and
$s \equiv 1/\log (L/L_0)$  
\begin{equation}
g(L) = \frac{\sqrt{3}k}{2} s \left( 1 + s \frac{k}{2} \log s  +
s^2 \left( \frac{k^2}{4} \log^2 s  
- \frac{k(k+10)}{24}\right) + \cdots
\right).
\label{run}
\end{equation}
Substituting this running coupling constant into
the $c$-function (\ref{cfun}), we obtain 
\begin{eqnarray}
c(g(L)) &=& c_0 + \frac{A_1}{\log^3 L/L_0} +
\frac{A_2 + A_2 ' \log \log L/L_0}{\log ^4 L/L_0} \\ \nonumber
&+& \frac{A_3 + A_3 ' \log \log L/L_0 +
A_3'' (\log \log L/L_0)^2}{\log ^5 L/L_0} + {\rm O}(\log^{-6} L/L_0),
\label{clog}
\end{eqnarray}
where the central charge $c_0$ of level $k$
$su(2)$ WZW model and other coefficients are
\begin{eqnarray}
c_0 &=& \frac{3k}{(k+2)}, \\ \nonumber 
A_1 &=& \frac{3}{8} k^2, \\ \nonumber
A_2 &=& \frac{9}{64} k^3, \ \ \
A_2 '= -\frac{9}{16} k^3, \\ \nonumber
A_3 &=& -\frac{9}{16} k^3, \ \ \
A_3 ' = -\frac{9}{32} k^4, \ \ \
A_3'' = \frac{9}{16} k^4.  
\label{cfclog}
\end{eqnarray}
This formula has only one fitting parameter $L_0$. 
The logarithmic finite size correction to the ground state
energy $E_0(L)$ in the Faddeev-Takhtajan chain with
finite length $L$ is calculated by
\begin{equation}
E_0(L) = e_\infty L - \frac{\pi \hbar v}{6 L} c(g(L))
\end{equation}
with  the formula (\ref{clog}), where $e_\infty$
is energy density in the infinite length limit and $v$ is
the spin wave velocity.
This formula (\ref{clog}) fits well the data of the ground state energy
in a numerical
Bethe ansatz obtained by Nomura \cite{N} in $k=1$ case. 
He shows the row data of low lying energy levels in the Heisenberg 
antiferromagnetic chain with $L=$ 256 - 16384 lattice sites
with unit lattice spacing.
To check the consistency with
all data including the excited energy, we have to
consider the gamma functions of two primary fields in WZW model.
In the level one $su(2)$ WZW model,
we have four primary fields which 
correspond a singlet and a triplet excitations
with conformal dimension (1/4, 1/4). We write their gamma function
\begin{eqnarray}
\gamma^{t}(g) &=& \gamma^t _0 + \gamma^t _1 b_0 g + \gamma^t _2 (b_0 g)^2
+\gamma^t _3 (b_0 g)^3 + {\rm O}(g^4), \\ \nonumber
\gamma^{s}(g) &=& \gamma^s _0 + \gamma^s _1 b_0 g + \gamma^s _2 (b_0 g)^2
+\gamma^s _3 (b_0 g)^3 + {\rm O}(g^4).
\label{gammas}
\end{eqnarray}
The level one $su(2)$ WZW model gives
$$
\gamma^t _0 = \gamma^s _0 = 1/2,
$$
and one loop RG calculation gives
$$\gamma^t _1 = -1/2, \ \ \ \gamma^s _1 = 3/2, \ \ \  
b_0 = \frac{4}{\sqrt{3}},
$$
however other coefficients have never been calculated. 
Note that the 
first excited state  
has three fold degeneracy because of the $su(2)$ symmetry. 
First, we fit the data at two loop level.
In this order, the invariant scale is given by $L_0 ^{-1} = 1.74087$ 
for fitting the data of the ground state energy in \cite{N}
by the formula (\ref{clog}).
Unfortunately in the present stage, we have only
the first order gamma function of the primary fields. 
Though we cannot check his data of excited 
energy in second order explicitly,
we can fix the coefficients $\gamma^t _2$ and $\gamma^s _2$
by fitting the data as follows:
\begin{eqnarray}
\gamma^{t}_{2Loop}(g) &=& 1.00012 \left(\frac{1}{2} - 
\frac{1}{2} b_0 g \right)
-0.0619577 (b_0 g)^2 , \\ \nonumber
\gamma^{s}_{2Loop}(g) &=& 0.999499 \left( \frac{1}{2} + 
\frac{3}{2} b_0 g \right)
+ 1.00147 (b_0 g)^2 .
\end{eqnarray}
Next, we discuss three loop order.
The invariant length scale becomes $L_0 ^{-1} = 2.14538$
to fit the ground state energy in this order. 
The gamma functions for fitting the excited energy become
\begin{eqnarray}
\gamma^{t}_{3Loop}(g) &=& 0.999868 \left(\frac{1}{2}-\frac{1}{2} b_0 g \right)
-0.00772577 (b_0 g)^2 - 0.284163 (b_0 g)^3, \\ \nonumber
\gamma^{s}_{3Loop}(g) &=& 0.999932 \left( \frac{1}{2} + \frac{3}{2} b_0 g \right)
+ 0.863347 (b_0 g)^2 + 0.64291(b_0 g)^3. 
\end{eqnarray}
It is important that both coefficients of the first order 
are quite close to unity as in two loop order. 
This fact shows a good agreement between our three loop calculation and 
the numerical Bethe ansatz. 
On the other hand, other coefficients change
from those in the two loop order. 

In this fitting, we compare our calculation of some typical values
with those by Nomura in the numerical Bethe ansatz 
\begin{eqnarray}
c_{Bethe}(g(256)) &=& 1.00103233, \ \ \  
c_{Bethe}(g(16384)) = 1.000239164, \\ \nonumber
c_{2Loop}(g(256)) &=& 1.00101926, \ \ \  
c_{2Loop}(g(16384)) = 1.000241815, \\ \nonumber
c_{3Loop}(g(256)) &=& 1.00101115, \ \ \  
c_{3Loop}(g(16384)) = 1.000243613, 
\end{eqnarray}
\begin{eqnarray}
\gamma^{t}_{Bethe}(g(256)) &=& 0.464830, \ \ \  
\gamma^{t}_{Bethe}(g(16384)) = 0.478297, \\ \nonumber
\gamma^{t}_{2Loop}(g(256)) &=& 0.464835, \ \ \  
\gamma^{t}_{2Loop}(g(16384)) = 0.478301, \\ \nonumber
\gamma^{t}_{3Loop}(g(256)) &=& 0.464832, \ \ \  
\gamma^{t}_{3Loop}(g(16384)) = 0.478296, 
\end{eqnarray}
\begin{eqnarray}
\gamma^{s}_{Bethe}(g(256)) &=& 0.620490, \ \ \ 
\gamma^{s}_{Bethe}(g(16384)) = 0.570912, \\ \nonumber 
\gamma^{s}_{2Loop}(g(256)) &=& 0.620513, \ \ \ 
\gamma^{s}_{2Loop}(g(16384)) = 0.570916, \\ \nonumber 
\gamma^{s}_{3Loop}(g(256)) &=& 0.620493, \ \ \ 
\gamma^{s}_{3Loop}(g(16384)) = 0.570910,  
 \end{eqnarray}
The two loop calculation is slightly better than the 
three loop one with respect to the $c$-function.
It is not less trivial that
this three loop gamma functions fit the excited energy levels very well,
since this fitting has two more parameters 
$\gamma^t _3, \gamma^s _3$.
These results of the RG calculation seem 
consistent with the numerical Bethe ansatz.

Here, we discuss consistency of our results with
those obtained by several authors \cite{N,KM,Klmp,WE}.
First, we comment on Nomura's fitting of 
his numerical data based on 
a two loop RG calculation \cite{N}, which gives $A_1=0.365162$.
He take into account only first order of
corrections to the $c$-function, 
although he uses the running coupling constant 
obtained from the two loop beta function.
This is consistent way to
take into account $\log \log L/L_0 \log^{-4} L/L_0$
term in the $c$-function (\ref{clog}). 
However, we cannot neglect $\log ^{-4} L/L_0$ term 
to fit the numerical data, since  $\log ^{-4} L/L_0$ term
behaves as large as $\log \log L/L_0 \log^{-4} L/L_0$
in numerical data of the model with 
not so large degrees of freedom (256-16384 sites).
To obtain more accurate fitting, 
we should take into account $g^4$ order in $c$-function
and $g^2$ order in the gamma function of the primary fields.
He fixes the parameter $L_0 ^{-1}  = 0.56532$ 
in such a way that the one loop correction to the triplet 
excitation energy with the two loop running coupling 
constant fits the numerical data of the
triplet excitation. 
It is possible to fit only one observable 
by adjusting the parameter $L_0$
with neglecting the second order correction in it,
since the second order logarithmic correction
can be absorbed into the first order one 
by adjusting the parameter $L_0$. In this procedure however,
the second order logarithmic correction cannot be absorbed 
into first order one
to fit observables more than 
one without the second order correction. 
Despite his recognition of the $g^2$ order correction 
to the singlet excitation energy and 
$L^{-2}$ correction to the ground state energy by
the irrelevant operator $L_{-2} \bar{L}_{-2} {\bf 1}$,
his obtained $c$-function is
$c(g(L))= 1 + 0.36516 \left( \frac{b_0 g}{2} \right)^3 + \frac{1.666}{L^2},$
which gives not so good agreement with the leading coefficient 
$A_1= 0.375$. 
If we take care of $\log^{-4} L/L_0$ term from 
the $g^4$ term in the $c$-function
and $\log ^{-2} L/L_0 $ term from the
$g^2$ term in the gamma function $\gamma^t(g)$,
the numerical data should be 
fitted with sufficient accuracy even in two loop order. 
Actually, Karbach and M\"{u}tter obtain $A_1=0.375$
$A_2 '= -0.73$ and $A_2=0.15$ with recognition of 
$\log^{-4} L/L_0$ term in the $c$-function with 
a careful treatment of fitting their data in numerical Bethe ansatz.
Their obtained $A_1$ agrees with ours, and also  
Kl\"{u}mper shows this result $A_1=0.375$ 
by calculating the specific heat
in his new method of solving 
a set of non-linear equations based on a quantum transfer
matrix for the thermodymamic quantities partially using 
a numerical calculation. 
We will see the relation between the specific heat
and the ground state energy in the next section.
Next, we discuss the two loop order coefficients. Our results 
$A_2 '= -9/16 = -0.5625$ and 
$A_2 = 9/64 + 9/8 \log L_0 \sim -0.483058$, with the 
two loop fitting parameter $L_0 ^{-1} =1.74087$ in (\ref{clog}) 
does not agree 
with those by Karbach and M\"{u}tter. 
We expect our formula with the three loop correction
(\ref{clog}) will be useful to fit their data, 
eventhough fitting the numerical
data for the higher order coefficients seems extremly difficult.
 
Finally, we mention the results obtained by Woynarovich and Eckel
with Euler-MacLaurin expansion in analytic Bethe ansatz \cite{WE} . 
The coefficients $\gamma^t _0$, $\gamma^t _1$, $\gamma^s_0$ and 
$\gamma^s _1$ in excited energy level and $c_0$ in the WZW
model agree with those obtained by Euler-MacLaurin 
expansion in the 
Bethe ansatz. On the other hand, the first coefficient 
in the logarithmic correction in eq.(\ref{cfclog})
is calculated as
$
A_1
= 0.343347 \cdots,
$ in \cite{WE}
which does not agree with
the result $A_1 = 3/8 = 0.375$ obtained by the
WZW model. This is well-known contradiction
pointed out in \cite{N,KM,Klmp,A}. Since our fitting of numerical Bethe ansatz,
an appropriate treatment of the numerical Bethe ansatz \cite{KM} and 
Kl\"{u}mper's result \cite{Klmp} support 
the WZW model, we believe $A_1 = 3/8$. 

\section{Specific heat and suceptibility}

Next, we calculate the specific heat per unit length.
The partition function of one
dimensional
quantum system is given by
\begin{equation}
Z(L, M) = \tr e^{- M H/{\hbar v} }.
\end{equation}
$M= \frac{\hbar v}{k_{\rm B} T}$ 
has the dimension of length, where $v$ is a spin wave velocity, 
$k_{\rm B}$ is the Boltzmann constant and $T$ is the temperature.
The partition function possesses a modular invariance
\begin{equation}
Z(L, M) = Z(M, L),
\end{equation} 
and the free energy per unit length
$f(L, M) = -\frac{k_{\rm B}T}{L} \log Z(L, M)$ possesses it as
well.
In the low temperature limit $M \rightarrow \infty$, 
the free energy becomes
\begin{eqnarray}
f(L, \infty) &=& E_0(L)/L 
\\ \nonumber 
&=& e_\infty - \frac{\pi \hbar v}{6 L^2} c(g(L)). 
\end{eqnarray}
The modular invariance gives the temperature dependence of 
the free energy per unit length
for the infinitely long chain as follows:
\begin{eqnarray}
f(\infty, M) &=& f(M, \infty) \\ \nonumber
&=& e_\infty - \frac{\pi \hbar v}{6 M^2} c(g(M)).
\end{eqnarray}
The specific heat per unit length
$C(T) = - T \frac{\partial^2 f}{\partial T^2}$
is calculated in a low temperature expansion 
\begin{eqnarray}
C(T) &=& \frac{\pi k_{\rm B} ^2 T}{3 \hbar v} 
( c_0 + \frac{B_1}{\log^3 T_0/T}
+ \frac{B_2 + B_2' \log \log T_0/T }{\log^4 T_0/T} \\ \nonumber
&+& \frac{B_3+ B_3' \log \log T_0/T+ B_3'' 
(\log \log T_0/T)^2}{\log^5 T_0/T}
+ {\rm O}(\log^{-6} T_0/T) \ ),
\label{specific}
\end{eqnarray}
where $T_0 = \frac{\hbar v}{k_{\rm B} L_0}$ 
is a parameter with the dimension of temperature.
The coefficients are given by
\begin{eqnarray}
c_0 &=& \frac{3k}{k+2} \\ \nonumber
B_1 &=& \frac{3}{8} k^2, \\ \nonumber
B_2 &=& \frac{9}{64} k^2 (k + 4), \ \ \
B_2'= - \frac{9}{16} k^3, \\ \nonumber
B_3 &=& -\frac{9}{16} k^2 (k - 4), \ \ \
B_3'= -\frac{9}{32} k^3 (k + 4), \ \ \
B_3''= \frac{9}{16} k^4.
\end{eqnarray}
This formula shows $A_1=B_1$, $A_2 ' = B_2 '$ 
and $A_3 ''= B_3 ''$ given in eq.(\ref{clog}), and therefore 
our result is consistent with 
that obtained by Kl\"{u}mper for the specific heat
\cite{Klmp}. 
The specific heat per unit length of 
any model in the universality class 
including $S=1/2$ Heisenberg antiferromagnetic chain
obeys this result with $k=1$.
For example, the specific heat per unit length 
of $S=1/2$ antiferromagnetic 
chain with a small perturbation of a second neighbor interaction
satisfies the same formula by adjusting $T_0$. 
This shift occurs through the change of
the initial coupling constant $g(a)$.
As proposed in \cite{K},
this low temperature behavior of the 
specific heat per unit length (\ref{specific})
in the $S=1/2 \ ( k=1 )$ Heisenberg chain  
will have to be observed experimentally, as well as its suceptibility.
Eggert, Affleck and Takahashi
calculated its suceptibility
by one loop renormalization group and the 
Bethe ansatz for the Heisenberg chain \cite{EAT}. 
The result has been 
observed by N. Motoyama, H. Eisaki and S. Uchida, experimentally
in an antiferromagnetic chain 
${\rm Sr}_2{\rm Cu O}_3$ \cite{U}.
Here we discuss the data fitting of both 
the suceptibility and the specific heat which are observed. 
The Heisenberg hamiltonian is 
$$
H= J \sum_i {\mbox{\boldmath S}}_i \cdot {\mbox{\boldmath S}}_{i+1}.
$$
The spin wave velocity is given by 
$$
v= \pi J/2,
$$
which is determined from the Bethe ansatz.
Eggert {\it et al.} gives
the best fit $T_0 = 7.7 J$  of the one loop renormalization group 
calculation for the suceptibility by 
the Bethe ansatz calculation of the Heisenberg chain 
\cite{EAT}.
They argued that the $\log^{-2} T_0/T$ term in the suceptibility
can be eliminated by shifting $T_0$.
In addition to the $\log ^{-2} T_0/T$ term
however, we have further logarithmic corrections
from the two loop running coupling constant (\ref{run}) 
in the one loop corrected suceptibility
$$
\chi(T) = \frac{1}{2 \pi v} + \frac{1}{4 \pi v \log T_0/T} 
\left( 1- \frac{\log \log T_0/T}{2 \log T_0/T}  \right)
+ {\rm O}(\log^{-3} T_0/T).
$$
This formula should be better for fitting the suceptibility
calculated by the Bethe ansatz.
To fit the specific heat at the same time,
however, the adjusted $T_0$ for the suceptibility
cannot be used for the specific heat with the same accuracy level
as the suceptibility. This is because
the same reson as in the fitting of numerical Bethe ansatz
for the logarithmic finite size correction we argued in the 
previous section. The elimination of both the 
$\log ^{-2} T_0/T$ term in the suceptibility
and $\log^{-4}T_0/T $ term in the specific heat 
cannot be done by shifting the only one 
parameter at the same time.
We have to calculate the coefficient of 
$\log ^{-2} T_0/T$ term  
in the suceptibility to fit 
both specific heat and the suceptibility.
To do so, we have to calculate the $g^2$ order
correction in the suceptibility explicitly. 

\section{Conclusion}

In this paper, we have shown a simple calculation method for
the higher order beta function of a perturbed CFT
by utilizing the constraints given by Zamolodchikov's $c$-theorem. 
In a particular CFT model $\widehat{su(2)_k} \oplus \widehat{su(2)_l }
/ \widehat{su(2)_{k+l}}$,
the known exact form of the OPE coefficient of the slightly relevant operator
enables us to calculate the beta function up to three loop order
in the $\varepsilon$-expansion, where $\varepsilon \equiv 4/(k+l+2)$.
In the limit $\varepsilon \rightarrow 0$
as $l \rightarrow \infty$, we have the level $k$ WZW model
with a marginal perturbation. The obtained three loop beta
function is useful to study a quantum spin chain model described 
in the marginally perturbed WZW model.
The logarithmic finite size correction 
to the ground state energy in the $su(2)$
quantum spin chain model has been calculated up to the $\log^{-5} L$
order as well as the logarithmic temperature dependence of the specific 
heat. The obtained formula of finite size correction fits the data of a
numerical Bethe ansatz quite well. The low temperature behavior of the 
specific heat obtained here should be observed experimentally.
We provide a slightly better fitting function of the suceptibility
than the one loop renormalization group calculation.
Completely consistent treatment of both 
the suceptibility and the specific heat
will be reported soon.
\\

{\large {\bf Acknowlegments}} 

The author would like to thank M.-H. Kato 
for explaining his recent work and 
enlightening discussions. He thanks Y. Hatsugai
for helpful comments on experimental physics. He is grateful
to M. Itoi for helpful work of the data fitting.

{\large {\bf Note added}}
After writening this paper, the author discussed with N. Motoyama
who explained that 
the spin specific heat of an antiferromagnetic chain 
${\rm Sr}_2{\rm Cu O}_3$ \cite{U} cannot be 
separated from phonon specific heat with extremely large value
compared to the spin part.



\begin{thebibliography}{10}
\bibitem{LG} A. B. Zamolodchikov, Sov. J. Nucl. Phys. {\bf 44}
(1986) 529; \\ D. A. Kastor, E. J. Martinec and S. H. Shenker,
Nucl. Phys. B{\bf 316} (1989) 590.
\bibitem{Z} A. B. Zamolodchikov, Sov. J. Nucl. Phys. {\bf 46}
(1987) 1090; \\
A. W. W. Ludwig and J. L. Cardy, Nucl. Phys. {\bf B 285}(1987)687
\bibitem{K} M. -H. Kato, ``{\it Renormalization group
flows and crossover scaling functions in deformed coset models}",  
Nihon university Ph.D. thesis, (1998).
\bibitem{N}
K. Nomura, Phys. Rev. B{\bf 48} (1993) 16814.
\bibitem{U}
N. Motoyama, H. Eisaki and S. Uchida,
Phys.Rev.Lett. {\bf 76} (1996)1312.
\bibitem{AlZ}
Al. B. Zamolodchikov, Nucl. Phys. B{\bf 366}(1991)122.
\bibitem{ABL}
C. Ahn, D. Bernard and A. LeClair, Nucl. Phys. B{\bf 346}(1990)409.
\bibitem{CSS}
\v{C}. Crnkovi\'{c}, G. M. Sotkov and M. Stanishkov, Phys. Lett.
{\bf B226} (1989) 297.
\bibitem{KM}
M. Karbach and K.-H. M\"{u}tter, J. Phys. A {\bf 28} (1995) 4469.
\bibitem{Klmp}
Kl\"{u}mper, ``{\it The spin-1/2 Heisenberg chain: thermodynamics,
quantum criticality and spin-Peierls exponents}", cond-mat/9803225.
\bibitem{WE}
F. Woynarovich and H. -P. Eckel, J. Phys. A{\bf 20}
(1987) L97
; A{\bf 20}L443.
\bibitem{A} 
I. Affleck, D. Gepner, H. J. Schulz and T. Zimann,
J. Phys. A{\bf 22} (1989) 511.
\bibitem{EAT}
S. Eggert, I. Affleck and M. Takahashi,
Phys. Rev. Lett. {\bf 73}(1994)332
\end{thebibliography}
\end{document}